From burigana@tesre.bo.cnr.it Tue Mar  9 13:59:43 1999
Date: Tue, 9 Mar 1999 11:54:26 +0100
From: Carlo Burigana <burigana@tesre.bo.cnr.it>
To: toffol@pinon.ccu.uniovi.es, toffol@tesre.bo.cnr.it
Subject: santander98_buriganaetal_accepted.tex ; puoi metterlo nel babbage?

\def   \ni {\noindent}

\def   \ssk {\vskip  5truept}

\def   \bsk {\vskip 15truept}
 
\def   \newpage {\vfill\eject}
\def   \newline {\hfil\break}

\def\and{{\&}}

\def\dex1{\mbox{dex}}
\def\dex{\hbox{\rm dex}}

\def\eg1{{e.g.\/}}

\def\gp{\hbox{\rlap{\hbox{.}}\raise 5.truept \hbox{{\small$\circ$}}}}
\def\gradip{\hbox{\rlap{\hbox{.}}\raise 5.truept \hbox{{\small $\circ$}}}}
\def\orep{\hbox{\rlap{\hbox{.}}\raise 5.truept \hbox{{\small $h$}}}}

\def\magcir{\ \raise-2.truept\hbox{\rlap{\hbox{$\sim$}}\raise5.truept
\hbox{$>$}\ }}

\def\mincir{\ \raise-2.truept\hbox{\rlap{\hbox{$\sim$}}\raise5.truept
\hbox{$<$}\ }}

\def\underline{}

\def\lsim{\,\lower2truept\hbox{${< \atop\hbox{\raise4truept\hbox{$\sim$}}}$}\,}
\def\gsim{\,\lower2truept\hbox{${> \atop\hbox{\raise4truept\hbox{$\sim$}}}$}\,}

\def\s{\,{\rm s}}

\def\s_fit{\mbox{$\sigma_{\rm bf}$}}

\def\dex{\mbox{dex}}

\def\eg{{\it e.g}}

\def\$\sigma$c{\mbox{$\$\sigma$_c$}}

\def\gp{\hbox{\rlap{\hbox{.}}\raise 5.truept \hbox{{\small$\circ$}}}}

\def\ni{\noindent}                          

\def\lsim{\, \lower2truept\hbox{${< \atop\hbox{\raise4truept\hbox{$\sim$}}}$}\,}
\def\gsim{\, \lower2truept\hbox{${> \atop\hbox{\raise4truept\hbox{$\sim$}}}$}\,}
\documentstyle[epsfig]{article}
\begin{document}

%
\def\la{\mathrel{\mathchoice {\vcenter{\offinterlineskip\halign{\hfil
$\displaystyle##$\hfil\cr<\cr\sim\cr}}}
{\vcenter{\offinterlineskip\halign{\hfil$\textstyle##$\hfil\cr
<\cr\sim\cr}}}
{\vcenter{\offinterlineskip\halign{\hfil$\scriptstyle##$\hfil\cr
<\cr\sim\cr}}}
{\vcenter{\offinterlineskip\halign{\hfil$\scriptscriptstyle##$\hfil\cr
<\cr\sim\cr}}}}}
\def\ga{\mathrel{\mathchoice {\vcenter{\offinterlineskip\halign{\hfil
$\displaystyle##$\hfil\cr>\cr\sim\cr}}}
{\vcenter{\offinterlineskip\halign{\hfil$\textstyle##$\hfil\cr
>\cr\sim\cr}}}
{\vcenter{\offinterlineskip\halign{\hfil$\scriptstyle##$\hfil\cr
>\cr\sim\cr}}}
{\vcenter{\offinterlineskip\halign{\hfil$\scriptscriptstyle##$\hfil\cr
>\cr\sim\cr}}}}}
\def\degr{\hbox{$^\circ$}}
\def\arcmin{\hbox{$^\prime$}}
\def\arcsec{\hbox{$^{\prime\prime}$}}

\hsize 5truein
\vsize 8truein
\font\abstract=cmr8
\font\keywords=cmr8
\font\caption=cmr8
\font\references=cmr8
\font\text=cmr10
\font\affiliation=cmssi10
\font\author=cmss10
\font\mc=cmss8
\font\title=cmssbx10 scaled\magstep2
\font\alcit=cmti7 scaled\magstephalf
\font\alcin=cmr6 
\font\ita=cmti8
\font\mma=cmr8
\def\ref{\par\noindent\hangindent 15pt}
\null


\title{\ni 
On Planck simulations: towards ``2-nd order'' analyses}

\bsk 
\author{\ni C.~Burigana$^1$, D.~Maino$^2$, N.~Mandolesi$^1$, F.~Villa$^1$, 
L.~Valenziano$^1$, M.~Bersanelli$^3$, L.~Danese$^2$, 
L.~Toffolatti$^{4,5}$ \& F.~Arg\"ueso$^6$}
\bsk
\affiliation{$^1$Istituto TeSRE, CNR, Bologna, Italy; 
$^2$SISSA -- International School for Advanced Studies, Trieste, Italy;
$^3$IFC, CNR, Milano, Italy;
$^4$Osservatorio Astronomico di Padova, Italy;
$^5$Departamento de F\'\i{sica}, Universidad de Oviedo, Spain;
$^6$Departamento de Matem\'aticas, Universidad de Oviedo, Spain.}
\bsk
\baselineskip = 9pt

\abstract{ABSTRACT -- \ni
We simulate Planck observations by adopting a detailed model of the
microwave sky including monopole, dipole, anisotropies of
the cosmic microwave background (CMB) and 
galactic and extragalactic foregrounds.
We estimate the impact of main beam optical aberrations  
on CMB anisotropy measurements in presence of  
extragalactic source fluctuations and we discuss the main implications 
for the Planck telescope design. By analysing the dipole pattern,
we quantify the Planck performance in the 
determination of CMB spectral distortion parameters in presence 
of foreground contaminations. 
}                                                    
\bsk
\baselineskip = 9pt

\bsk
\baselineskip = 12pt


\text{\ni 1. INTRODUCTION
\ssk
\ni     

The Planck satellite (Mandolesi et~al.~1998a; Puget et~al.~1998)
will map CMB anisotropies
over the entire sky at frequencies $\simeq 30 \div 857$~GHz with 
angular resolution of $\sim$~30$'\div$5$'$ and 
sensitivities of $\sim 5 \div 15 \mu$K per resolution element. 
It will lead to a determination of the
primordial angular power spectrum $C_l$ up to multipoles 
$l$$\sim$$(1 \div 2) \times 10^3$
and of the fundamental cosmological parameters
with unprecedented accuracy.

To reach the mission goals all 
instrumental and astrophysical systematic effects must
be carefully controlled. 
We have simulated Planck observations by adopting a detailed model 
of the  microwave sky -- including CMB monopole, dipole and fluctuations
together with galactic and extragalactic foregrounds (Section 2) --
to study two different kinds of ``$2-nd$ order'' features 
in which astrophysical and/or  
instrumental effects are treated in combination.

The contamination from galactic diffuse emission and 
extragalactic source fluctuations,
widely discussed in the literature
(see, e.g., Lasenby 1997; Toffolatti et~al.~1998; De~Zotti \& Toffolatti 1998),
depends on the frequency and angular scale. 
A more refined analysis requires coupling the sky signal with
realistic instrument response.
At Low Frequency Instrument (LFI) channels,  
main beam distortions introduce, 
in the case of a pure CMB anisotropy sky, an added (non--white) noise 
of order of $\sim$~2$\div$4~$\mu$K, depending on 
the level of optical distortions and beam width, 
and also on the $C_l$'s normalization and the cosmological model 
(Burigana et~al.~1998; Mandolesi et~al.~1998b).
In the 30 GHz channel this effect can be $\sim$~3 times larger at 
low galactic latitudes, 
because of the Galaxy emission gradients (Burigana et~al. 

\newpage

\noindent
1998).
In Section 3 we evaluate the impact on anisotropy measurements of 
main beam distortions in presence of  
extragalactic source fluctuations 
and discuss the main implications for the Planck telescope design. 

The accurate knowledge of the microwave sky dipole pattern can be used to
derive detailed information on the local system velocity with
respect to the CMB frame (Fixsen et~al.~1996)
and to constrain
possible CMB spectral distortions
(see, e.g., Burigana et~al.~1995),
that yield a dipole amplitude $\Delta T_D$,
sensitive to the CMB spectrum derivative
(Danese \& De Zotti 
1981).
This method has the advantage of 
circumventing the problem of high accuracy absolute calibration needed 
in traditional CMB spectrum experiments. 
In Section 4 we estimate the Planck performance in the 
determination of CMB spectral
distortion parameters from the analysis of the dipole pattern
in presence of foreground contaminations.



\bsk
\ni 2. GENERATION OF A SIMULATED MICROWAVE SKY
\ssk
\ni 

$\bullet$ {\it Modelling the CMB pattern} -- 
The CMB monopole and dipole,
for blackbody or distorted CMB spectra,
have been generated by using the Lorentz
invariance of photon distribution functions, $\eta$, in the phase space 
(Compton--Getting effect):
$\eta _{obs} (\nu_{obs},\vec n) = \eta _{CMB} (\nu _{CMB}) \,$,
where $\nu _{obs}$ is the observation frequency, 
$\nu _{CMB} = \nu _{obs} (1+\vec \beta \times \vec n) / \sqrt{1+\beta^2}$
is the corresponding frequency in the CMB rest frame, $\vec n$ is the unit vector
of the photon propagation direction and 
$\vec \beta = \vec v/c$ the observer velocity. 
For gaussian models, the CMB anisotropies at $l \ge 2$ can be simulated 
by following the standard spherical harmonic expansion (see, e.g., 
Burigana et~al.~1998) or by using FFT techniques which
take advantage of equatorial pixelisations (Muciaccia et~al.~1997).


$\bullet$ {\it Modelling the Galaxy emission} -- 
The Haslam map at 408~MHz (Haslam et~al.~1982) is
the only full-sky map currently available albeit 
large sky areas are sampled at 
1420~MHz (Reich \& Reich 1986) and at 2300~MHz (Jonas et~al.~1998). 
To clean these maps from free-free emission 
we use
a 2.7~GHz compilation of $\sim$ 7000 HII sources (C.~Witebsky et~al.~1978,
private communication) 
at resolution of $\sim$~1$^{\circ}$.
We use a spectral index $\beta_{ff} = 2.1$ from 2.7 to 1~GHz 
and $\beta_{ff} = 0$ below 1~GHz.
We then combine the synchrotron  maps producing a spectral 
index map between
408-2300~MHz with a resolution of $\lsim 2^{\circ} \div 3^{\circ}$ 
($<\beta_{sync}> \sim 2.8$). 
This spectral index map is used to scale the synchrotron  component
down to $\sim$ 10~GHz. In fact,  for 
typical (local) values of the galactic magnetic field ($\sim 2.5\mu$G), the knee
in the electron energy spectrum in cosmic rays ($\sim$ 15 Gev) 
corresponds to $\sim$~10~GHz (Platania et~al.~1998). 
>From the synchrotron map obtained at 10~GHz and 
the DMR 31.5~GHz map we derive a high
frequency spectral index map for scaling the synchrotron  component
up to Planck frequencies. 
These maps have a poor resolution and the synchrotron structure needs 
to be extrapolated to Planck angular scales.
An estimate of the synchrotron angular power spectrum
and of its spectral index, $\gamma$ ($C_l \propto l^{-\gamma}$), 
has been provided by Lasenby (1997); we used $\gamma = 3$ 
for the angular structure extrapolation (Burigana et~al.~1998).
Schlegel et~al.~(1998) provided a map of dust emission at 100$\mu$m merging
the DIRBE and IRAS results to produce a map with IRAS resolution ($\simeq 7'$) 
but with DIRBE calibration quality. They also provided a map of dust
temperature, $T_d$, by adopting a modified blackbody emissivity law, 
$I_\nu \propto B_\nu(T_d) \nu^{\alpha}$, with $\alpha =2$. This can be used to scale the dust
emission map to Planck frequencies 
using the dust temperature map as input for the
$B_\nu(T_d)$ function. Unfortunately the dust temperature map has a
resolution of $\simeq 1^{\circ}$; 
again, we use an angular power spectrum $C_l \propto l^{-3}$ 
to scale the dust skies to the Planck proper resolution.
Merging maps at different frequencies with different instrumental features and
potential systematics may introduce some internal inconsistencies.
More data on diffuse galactic emission, particularly at low frequency, would
be extremely important.


$\bullet$ {\it Modelling the extragalactic source fluctuations} -- 
The simulated maps of point sources have been created
by an all--sky Poisson distribution of the known populations of extragalactic
sources in the $10^{-5}< S(\nu)< 10$ Jy flux range
exploiting the number counts of Toffolatti et al. (1998)
and neglecting the effect of clustering of sources.
The number counts have been calculated by adopting the Danese
et al. (1987) evolution model of
radio selected sources
and an average spectral index $\alpha=0$
for compact sources up to $\simeq 10^3$ GHz and a break to $\alpha=0.7$
at higher frequencies (see Impey \& Neugebauer 1988; 
De~Zotti \& Toffolatti 1998), 
and by the model C of Franceschini et al. (1994)
updated as in Burigana et~al.~(1997), to account for the isotropic 
sub-mm component estimated by Puget et~al.~(1996) and Fixsen et~al.~(1996). 
At bright fluxes, far--IR selected sources
should dominate the number counts at High Frequency Instrument
(HFI) channels for $\nu \gsim 300$~GHz, 
whereas radio selected sources should dominate at lower 
frequencies (Toffolatti et~al.~1998). 
Moreover, the angular power spectra calculated by Toffolatti et~al.~(1998)
allows us to simulated 
gaussian extragalactic source fluctuation skies, by using the method
adopted for generating CMB gaussian fluctuations.

At 353 GHz we also exploited, for comparison,
the number counts of model ``E'' of Guiderdoni et~al.~(1998) 
which better accounts for the far--IR extragalactic background
recently detected by the COBE--DIRBE team (Hauser et~al.~1998),
although it overestimates
the level of the isotropic sub-mm component
as derived by Fixsen et~al.~(1998). 
Model ``E'' of Guiderdoni et~al.~(1998) is found to produce
a Poisson confusion noise
higher by a factor $\sim 1.5$ 
with respect to the prediction of Toffolatti et~al.~(1998) 
at the same frequency
(De Zotti \& Toffolatti 1998).


$\bullet$ {\it Modelling the observed signal} -- 
We produce full sky maps, $T_{sky}$,
by adding the antenna temperatures from CMB (with or without spectral 
distortions), Galaxy emission and extragalactic source fluctuations. 
The white noise depends on instrumental performances and on
the observed signal and changes in the sky
according to the Planck scanning strategy.
Planck will perform differential mesurements and not 
absolute temperature observations; 
we then represent the final observation
in a given $i$-$th$ pixel in the form 
$T_i = R_i (T_{sky,i} + N_i - T_{x,i}^r) \,$, where $N_i$ is the noise
(the $1/f$ noise can be reduced at levels smaller than
the white noise by efficient cooling
and with destriping 
algorithms),
$T_{x,i}^r$ is a reference
temperature
subtracted in the differential data
and $R_i$ is a constant which accounts for the calibration.
Of course, the uncertainty on $R_i$ 
and the pixel to pixel variation of $T_{x,i}^r$ have to be 
much smaller than the Planck nominal sensitivity.
Then, we generate the ``observed'' map assuming 
a constant value, $T_x^r$, of $T_{x,i}^r$ $\, \forall \, i \, $.
We note that possible constant small off-sets in 
$T_{x}^r$ could be in principle accepted, 
not compromising an accurate knowledge of the anisotropy pattern.
A possible isotropic astrophysical foreground 
relevant at Planck frequencies not covered by other experiments
(from $\simeq 1$cm to $\simeq 1$mm)
may be very difficult to subtract at a high level of accuracy, 
being orders of magnitudes smaller than the CMB monopole.
To be conservative, we can mimic this effect 
by analyzing the ``observed'' map by assuming 
$T_x^r$ to be somewhat different from that adopted in the map generation;
this can also mimic our uncertainty in estimating the foreground level
at higher frequencies.
We arbitrarily generate the ``observed'' map with $R_i=R=1$ 
$\, \forall \, i \, $;
again, by analyzing it with a different value of $R$, we can estimate the 
impact of a systematic error in the absolute calibration (for example, 
without using the dipole itself, 
the calibration with the planets achieves an accuracy of few percent;
we will assume a relative error of 5\% in the absolute calibration 
for numerical estimates).

\bsk
\ni 3. BEAM DISTORTIONS AND EXTRAGALACTIC SOURCE FLUCTUATIONS
\ssk
\ni 

Bright sources, for example above a 5$\sigma$ clipping threshold, 
contaminate a relatively small number of pixels and can be clearly identified
by carring out multifrequency observations. On the other hand, source fluctuations
could in principle contaminate Planck observations in a way difficult to predict
in presence of optical aberrations. 
We analyze here this effect for the 30 GHz channel, the most critical 
of the LFI for this effect. The main beam shape has been computed following
the method described in Mandolesi et~al.~(1998b) for two extreme Planck telescope 
apertures: 1.3m and 1.75m. 
In the present symmetric configuration for the Planck focal plane unit,
the two 30 GHz beams 
present specular shapes:
we can then study only one beam.
We convolve a CMB anisotropy map 
with the simulated beam and with a set of symmetric gaussian 
beams with $25' \le FWHM \le 40'$~; 
by comparing these simulated observations
for a suitable number of telescope positions in the sky,
we find the FWHM of that gaussian beam 
which gives measurements most similar to those obtained by our beam.
For the 1.3m and 1.75m telescope we find an ``equivalent'' FWHM of 38$'$.8 and 28$'$.7
respectively.
We can then convolve our extragalactic source fluctuation sky with the 
simulated beam and with the symmetric gaussian beam with the ``equivalent'' FWHM.
The $rms$ of the differences
of the temperatures ``measured'' in these two cases
provides an estimate of the noise added by optical distortions. 
For the 1.3m and 1.75m telescopes we have a $rms$ value of 
$\sim 1.5 \mu$K and of
$\sim 0.8 \mu$K 
respectively.
This effect is significantly smaller than
the average final sensitivity and the analogous effect in presence of 
CMB anisotropy only and of Galaxy emission at low galactic latitudes. 
On the other hand
it is not negligible compared to the sensitivity at high ecliptic
latitudes, where the Planck integration time is much larger than the average. 
Increasing
the primary mirror size respect to the Phase A study (1.3m aperture),
necessary to achieve the key goal of 10$'$ resolution at 
the ``cosmological'' channel at 100~GHz,
helps also in reducing this kind of contamination
(the present baseline allows to significantly increase the primary mirror aperture).
Also, further improvements
in the optical design, like that suggested by our preliminary study 
on aplanatic configurations,
can hopefully reduce this effect.  
We stress the relevance of suppressing this effect by optimizing
the optical design,
being difficult and time consuming, if not impossible,
to reduce it in the data analysis only, 
in presence of many other kinds of systematic effects.

\bsk
\ni 4. CONSTRAINTS ON THE CMB SPECTRUM FROM THE DIPOLE PATTERN 
\ssk
\ni 

We fit our simulated maps by modelling only the CMB monopole and dipole
contributions to $T_{sky}$ and the local system velocity (we assume $T_0$ and 
$\vec \beta$ within the 95\% CL FIRAS limits, $a T_0^4$ being the present 
radiation energy density). 
We subtract the Galaxy emission, 
by using a model more rough than that adopted
for generating it, in order to test
the impact of a non accurate Galaxy subtraction on the 
recovering of the CMB distortion parameters.
We use the 408 MHz map (Haslam et~al.~1982) as a synchrotron template and 
scale it to 10~GHz with a spectral index $\beta_{sync} = 
2.8$; the dust emission ($\propto B_{\nu} (T_d) \nu^2$)
is scaled from the 100 $\mu$m map (Schlegel et~al.~1998) 
with a dust temperature $T_d = 20$K; 
$\beta_{sync}$ and $T_d$ are taken constant over the sky.
No extrapolation at small angular scales has been performed.
The difference between models with 
$\beta_{sync} = 2.65$ and 2.95 
and $T_d = 15$~K and 25~K respectively
for extrapolating synchrotron emission in the $\sim 1 \div 10$~GHz range 
and dust emission from $240 \mu$m provides an estimate
of the uncertainty of this Galaxy subtraction model. 
Conservatively, we did not subtract the contributions 
to small scales anisotropies from
primordial fluctuations and extragalactic sources, 
which give an additional uncertainty in this calculation.
On the contrary, we simply include in the 
quoted error in each pixel the variances of the input 
maps of CMB anisotropies 
and extragalactic source fluctuations, each being a scientific 
output of the Planck mission.

We have separately exploited three Planck channels (30, 100 and 353 GHz),
for a planckian spectrum and three kinds of ``standard'' distorted spectra:
the Bose--Einstein distortion, 
the comptonization distortion, described respectively by 
the chemical potential $\mu$ and the comptonization parameter
$u$, related to the fractional energy injected in the radiation field,
and the free--free distortion, described
by a parameter, $y_{B}$, related to the plasma thermal history at redshifts
less than $\approx 10^4$.

By fitting the dipole pattern in absence of foreground contaminations,
we recover exactly the input distortion parameters at each 
channel with very small statistical uncertainties
($\sim few \times 10^{-9}\div few \times 10^{-8}$ at 95\% CL). 
Similarly, gaussian extragalactic source
fluctuations insignificantly affect the input parameter recovering.
On the contrary, Poisson extragalactic
source fluctuations, Galaxy emission and CMB anisotropies 
at $l\ge 2$ degrade our capability of recovering the input 
distortion parameters (we have verified that an intrinsic CMB dipole 
with $C_1 \sim C_2$ does not affect our conclusions).
For example, by adopting an input planckian spectrum and working at 100~GHz
we obtain $\mu \sim 2.2 \times 10^{-5}$,
$u \sim 2.6 \times 10^{-6}$ and 
$y_B \sim -1.6 \times 10^{-5}$ with statistical uncertainties in the range 
$few \times 10^{-8} \div few \times 10^{-7}$.
In general,
working at low (high) frequencies is more advantageous for 
recovering $\mu$ and $y_B$ ($u$) distortions, as we have found from
the channels at 30 and 353~GHz.
By fitting the same observed maps by assuming a calibration constant $R=0.95$,
we find absolute differences between
the input distortion parameters and the recovered ones 
in the range $\sim 3 \times 10^{-6} \div 3 \times 10^{-5}$, quite close
to those previously obtained.
Finally, we consider also the presence of a possible ignored
foreground (we assume for the present estimates a value lower 
than the CMB monopole by a factor
$\sim 5\times 10^{-5}$ at 30 and 100 GHz and $\sim 10^{-4}$ at 353 GHz,
with no calibration error):
we find absolute differences between
input and recovered distortion parameters 
in the range $\sim 4 \times 10^{-6} \div 10^{-5}$, again with
very small statistical uncertainties. 
Then, the most important source of uncertainty derives from astrophysical
contaminations.
Similar degradations and uncertainties are obtained also for distorted spectra. 

Even in presence of all the uncertainties discussed above,
these preliminary results 
are promising, indicating that Planck could improve 
the present constraints on distortion parameters 
by a factor $\gsim 5$ 
[i.e. with errors 
$\Delta \mu \sim \Delta u \sim \Delta y_B \lsim (1 \div 2) \times 10^{-5}$]
or detect possible spectral distortions
with a similar level of uncertainty.

}

\bsk
\baselineskip = 9pt
{\abstract \ni ACKNOWLEDGMENTS

We gratefully acknowledge stimulating and helpful discussions with
G.~De~Zotti, P.~Platania and G.F.~Smoot.
We wish to thank B.~Guiderdoni for kindly providing us the number
counts of their model ``E''.
LT and FAG 
acknowledge partial financial support from 
the Spanish
Direcci\'on General de Ense\~nanza Superior (DGES), projects
PB95--0041 and PB95--1132--C02--02.}

\bsk
\baselineskip = 9pt


{\references \ni REFERENCES
\ssk













\ref
Burigana, C., De~Zotti, G., Danese, L.  1995, A\&A 303, 323

\ref
Burigana, C., et al. 1997, MNRAS 287, L17



\ref
Burigana, C., et al. 1998, A\&ASS 130, 551







\ref
Danese, L., De Zotti, G. 1981, A\&A 94, L33


\ref
Danese, L., et al. 1987, ApJ 318, L15





\ref
De Zotti, G., Toffolatti, L. 1998, this Conference


\ref
Fixsen, D.J., et al. 1996, ApJ 473, 576

\ref
Fixsen, D.J., et al. 1998, ApJ, in press (astro-ph/9803021)

\ref
Franceschini, A., et al. 1994, ApJ 427, 140






\ref
Guiderdoni, B., et al. 1998, MNRAS 295, 877

\ref
Haslam, C.G.T., et al. 1982, A\&ASS 47, 1

\ref
Hauser, M.G., et al. 1998, ApJ, in press (astro-ph/9806167)



\ref
Impey, C.D., Neugebauer, G. 1988, AJ 95, 307


\ref
Jonas, J.L., Bart, E.E., Nicolson, D. 1998, MNRAS, 297, 977








\ref
Lasenby, A.N. 1997, in proc. of {XVIth Moriond Astroph. Meeting}, 
pg. 453 (astro-ph/9611214)



\ref
Mandolesi, N., et al. 1998a, Planck LFI,
A Proposal Submitted to the ESA. 

\ref
Mandolesi, N., et al. 1998b, A\&A, submitted

\ref
Muciaccia, P.F., Natoli, P., Vittorio, N. 1997, ApJ 488, L63 


\ref
Platania, P., et al. 1998, ApJ 505, 473

\ref
Puget, J.-L., et al. 1996, A\&A 308, 5  

\ref
Puget, J.-L., et al. 1998, HFI for the Planck Mission,
A Proposal Submitted to the ESA.

\ref
Reich, P., Reich, W. 1986, A\&ASS 63, 205



\ref
Schlegel, D.J., Finkbeiner, D.P., Davis, M. 1998, ApJ 500, 525








\ref
Toffolatti, L., et al. 1998, MNRAS 297, 117







}                      

\end{document}